\documentstyle[12pt]{article}

\def\dfrac{\displaystyle\frac}
\def\eqlab#1{\label{eqn:#1}}
\def\eqref#1{Eq.(\ref{eqn:#1})}

\def\eqsref#1{Eqs.(\ref{eqn:#1})}

\def\eqvref#1{(\ref{eqn:#1})}

\def\mz{m_Z^{}}
\def\mh{m_h^{}}
\def\PR#1#2#3{Phys. Rev. {\bf #1} (#3) #2 }
\def\PRL#1#2#3{Phys. Rev. Lett. {\bf #1} (#3) #2 }
\def\PL#1#2#3{Phys. Lett. {\bf #1} (#3) #2 }

\def\NP#1#2#3{Nucl. Phys. {\bf #1} (#3) #2 }

\def\PTP#1#2#3{Prog. Theor. Phys. {\bf #1} (#3)#2 }
\def\EPJ#1#2#3{Eur. Phys. J. {\bf #1} (#3) #2 }
\begin{document}
\renewcommand{\thefootnote}{\fnsymbol{footnote}}
\renewcommand{\theenumi}{(\roman{enumi})}
\title{Are lepton flavor mixings in the democratic mass matrix
stable against quantum corrections?}
\author{
{N. Haba$^{1,2}$}\thanks{haba@eken.phys.nagoya-u.ac.jp}
{, Y. Matsui$^2$}\thanks{matsui@eken.phys.nagoya-u.ac.jp}
{, N. Okamura$^3$}\thanks{naotoshi.okamura@kek.jp}
{, and T. Suzuki$^2$}\thanks{tomoharu@eken.phys.nagoya-u.ac.jp}
\\
\\
\\
{\small \it $^1$Faculty of Engineering, Mie University,}
{\small \it Tsu Mie 514-8507, Japan}\\
{\small \it $^2$Department of Physics, Nagoya University,}
{\small \it Nagoya, 464-8602, Japan}\\
{\small \it $^3$Theory Group, KEK,
 Tsukuba Ibaraki 305-0801, Japan}}
\date{May 8, 2000}
\maketitle
\vspace{-12.5cm}
\begin{flushright}
hep-ph/0005064\\
DPNU-00-19
\\
KEK-TH-691\\
\end{flushright}
\vspace{10.5cm}
\vspace{-2.5cm}
\begin{center}
\end{center}
\renewcommand{\thefootnote}{\fnsymbol{footnote}}
%
%
\begin{abstract}

We investigate whether the lepton
flavor mixing angles in the so-called
democratic type of
mass matrix are stable against quantum corrections or not
in the minimal supersymmetric standard model with dimension five
operator which induces neutrino mass matrix.
By taking simple breaking patterns 
of $S_3{}_L \times S_3{}_R$ or
$O(3)_L \times O(3)_R$ flavor symmetries and
the scale where democratic textures are induced as
$O(10^{13})$ GeV,
we find that the stability of the lepton flavor
mixing angles in the democratic type of mass matrix
against quantum corrections
depends on the solar neutrino solutions.
The maximal flavor mixing of the vacuum oscillation
 solution is spoiled by
 quantum corrections in the experimental allowed region of 
 $\tan \beta$. 
The large angle MSW solution is spoiled by 
quantum corrections in the region of $\tan \beta > 10$.
The condition of
 $\tan \beta \leq 10$ is needed in order to
 obtain the suitable mass squared difference of
 the small angle MSW solution.
These strong constraints must be regarded
for the model building of the democratic
type of mass matrix.

\end{abstract}

\bigskip


\newpage

Recent neutrino oscillation experiments 
suggest the strong evidences of
tiny neutrino masses and lepton flavor mixings
\cite{solar4,Atm4,SK4,CHOOZ}.
Studies of the lepton flavor mixing matrix,
which is so-called Maki-Nakagawa-Sakata (MNS) matrix\cite{MNS},
will give us important cues of the physics
beyond the standard model.
One important study is finding the suitable
texture of quark and lepton mass matrices in order to search
the flavor symmetry existing behind.
The democratic type of mass matrix\cite{democratic}
is one of the most interesting
candidate of the texture of quark and lepton mass matrices,
since it can naturally explain
the reason why only masses of third generation particles
are large comparing to those of other generations.
This type of mass matrix can be derived 
by flavor symmetries of
$S_3{}_L \times S_3{}_R$\cite{S3a,S3b,demo-tani} or
$O(3)_L \times O(3)_R$\cite{O3}.
As for the neutrino sector,
it has been said that the
democratic type of mass matrix can induce the suitable
solutions of the atmospheric and the solar neutrino problems
\cite{S3a,S3b,demo-tani,O3}.
However, are lepton flavor mixing angles in the
democratic type of mass matrix
stable against quantum corrections?

In this paper,
we investigate whether the lepton
flavor mixing angles in the
democratic type of mass matrix are stable 
against quantum corrections or not
in the minimal supersymmetric standard model 
with the dimension five
operator which induces the neutrino Majorana mass matrix.
The superpotential of the lepton-Higgs interactions is
given by
\begin{equation}
{\cal W} =
y^{\rm e}_{ij} (H_{d} L_i) {E_{j}}
- \dfrac{1}{2} \kappa^{}_{ij} (H_{u} L_i)(H_{u} L_j) \,,
\label{superpot}
\end{equation}
where $\kappa$ is the coefficient of the dimension five operator,
and the indices $i,j$ $(=1 \sim 3)$
stand for the generation number.
$L_i$ and $E_{i}$ are chiral super-fields of $i$-th generation
lepton doublet and right-handed charged lepton, respectively.
$H_{u}$ ($H_{d}$) is the Higgs doublet
which gives Dirac masses to the up- (down-) type fermions.
We will show the  renormalization
group equation (RGE) analyses of the lepton flavor mixing angles
\cite{up2now,ELLNE1,HO1}.

We take the simple breaking patterns 
of $S_3{}_L \times S_3{}_R$ or
$O(3)_L \times O(3)_R$ symmetries, and
the scale where democratic textures are induced as
$O(10^{13})$ GeV.
Under the above conditions,
we find that the stability of the lepton flavor
mixing angles in the democratic type of mass matrix
against quantum corrections depends
on the solar neutrino solutions.
The maximal flavor mixing of the vacuum oscillation (VO)
solution\cite{Vacuum} is spoiled by
quantum corrections in the experimental allowed region of 
 $\tan \beta$. 
The value of $\tan \beta$ is the ratio between
the vacuum expectation values (VEVs) of the Higgs particles.
The large angle MSW (MSW-L)
solution\cite{MSW} is spoiled by 
quantum corrections in the region of $\tan \beta > 10$.
On the other hand, the condition of
$\tan \beta \leq 10$ is needed in order to
obtain the suitable mass squared difference of
the small angle MSW solution
(MSW-S)\cite{MSW}.
These strong constraints must be regarded
for the model building of the democratic
type of mass matrix.

\vspace{5mm}

At first, we discuss the democratic mass matrix,
which is based on $S_{3L} \times S_{3R}$ or
$O(3)_L \times O(3)_R$ flavor symmetries.
In the democratic type of mass matrix,
the charged lepton mass matrix is given by
\begin{equation}
M_{l} = \dfrac{c_l}{3}
\left(\begin{array}{ccc}
1 & 1 & 1 \\
1 & 1 & 1 \\
1 & 1 & 1
\end{array}\right)
+ M_l^{(c)}\,,
\eqlab{charged lepton}
\end{equation}
where $M_l^{(c)}$ includes flavor symmetry
breaking masses, which must be introduced to obtain
the suitable values of $m_e$ and $m_{\mu}$.
The matrix $M_l$ is diagonalized by the unitary matrix
$V_l=FL$ from the side of left-handed fields,
where
\begin{equation}
F=
\left(\begin{array}{ccc}
1/\sqrt{2} &  1/\sqrt{6} & 1/\sqrt{3} \\
-1/\sqrt{2} &  1/\sqrt{6} & 1/\sqrt{3} \\
          0 & -2/\sqrt{6} & 1/\sqrt{3}
\end{array}\right) .
\end{equation}
Since we do not know the definite structure of
$M_l^{(c)}$, we can not determine the
explicit form of the unitary matrix $L$.
Here we assume that off diagonal elements of
$L$ are
small as $L_{ij}\ll 1$ $(i \neq j)$
from the analogy of
the quark sector \cite{S3a,S3b,demo-tani,O3}.
Thus, we obtain the relation of $V_l \simeq F$,
which is used in the following discussions\footnote{
We will revive $L_{ij}$ in
the MSW-S solution later. 
}.

The neutrino mass matrix \footnote{
To obtain this mass matrix, 
there is an alternative way using $S_3$ symmetry \cite{Moha}
instead of $S_{3L} \times S_{3R}$ or $O(3)_L \times O(3)_R$.
 }
is given by
\begin{equation}
M_{\nu} = c_\nu \left\{
\left(\begin{array}{ccc}
1 & 0 & 0 \\
0 & 1 & 0 \\
0 & 0 & 1
\end{array}\right)
+ r
\left(\begin{array}{ccc}
1 & 1 & 1 \\
1 & 1 & 1 \\
1 & 1 & 1
\end{array}\right)
\right\}+ M_{\nu}^{(b)} \,,
\eqlab{nu-mass}
\end{equation}
since the neutrinos are Majorana particles.
In \eqref{nu-mass},
$c_\nu$ and $r$ can be taken as real and non-negative
 parameters,
 and we neglect $CP$ phases, for simplicity.
The mass matrix $M_{\nu}^{(b)}$ breaks flavor symmetries,
which must be introduced in order to
obtain the suitable mass squared differences
and mixing angles of neutrinos.
There are following two simple breaking patterns according to
the solar neutrino solutions. \\
(i): The simplest example of $M_{\nu}^{(b)}$ for
 the MSW-L and the VO solutions 
 is to introduce two real and non-negative
 parameters $\epsilon$ and $\delta$
$( 1 \gg \delta \gg \epsilon )$
in (2,2) and (3,3) elements in $M_{\nu}^{(b)}$,
where the neutrino mass matrix $M_{\nu}^{(i)}$ is given by
\begin{equation}
M_\nu^{(i)} = c_\nu
\left(\begin{array}{ccc}
1+r &   r & r \\
  r & 1+r+\epsilon & r \\
  r &   r & 1+r+\delta
\end{array}\right)\,.
\eqlab{after_break_i}
\end{equation}
\noindent
(ii): The simplest example of $M_{\nu}^{(b)}$ for
 the MSW-S solution 
 is to introduce two real and non-negative
 parameters $\epsilon$ and $\delta$
 $( 1 \gg \delta \gg \epsilon )$
 in (1,2), (2,1), and (3,3) elements in $M_{\nu}^{(b)}$,
 where the neutrino mass matrix $M_{\nu}^{(ii)}$ is given by
\begin{equation}
M_\nu^{(ii)} = c_\nu
\left(\begin{array}{ccc}
1+r &   r+\epsilon & r \\
  r+\epsilon & 1+r & r \\
  r &   r & 1+r+\delta
\end{array}\right)\,.
\eqlab{after_break_ii}
\end{equation}

When $r \gg \epsilon$, $\delta$,
the unitary matrix $U_{\nu}$,
which diagonalizes $M_\nu$,
becomes $F$ in both cases of (i) and (ii).
In this case the MNS matrix
approaches to the unit matrix as
\begin{equation}
\label{FF}
V_{MNS} = V_l^{\dagger} U_{\nu} \simeq L^{\dagger} F^{\dagger}F
\simeq L^{\dagger},
\end{equation}
which does not have any large mixing angles.
Therefore the magnitude of $r$ must be smaller than
$\epsilon$, $\delta$
in order to obtain the large flavor mixing of 
the atmospheric neutrino solution\footnote{
We do not consider the case of $r=-2/3$\cite{KK}
which gives degenerate neutrinos,
where one of mass eigenvalues has a opposite
sign from others.
It is because the simple symmetry breaking patterns
such as (i) and (ii) can not induce the large
lepton flavor mixing
as shown above.
In the case of $r \gg \epsilon , \delta$,
the flavor symmetry breaking textures must be complicated
in order to solve both the solar and the atmospheric
neutrino problems. }.
Thus, in the democratic type of mass matrix,
three neutrinos are degenerate with the same signs,
where
the RGE effects cannot be negligible as shown in case (c4)
in Ref.\cite{HO1}.
This is the reason why we need RGE analyses
 for the democratic type of mass matrix.

\noindent
Under the condition of $r \ll \epsilon , \delta$, 
 both simple breaking patterns of (i) and (ii)
 induce the large mixing angle of 
 $\sin^2 2 \theta_{23}\simeq 8/9$ 
 which is suitable for the atmospheric neutrino
 solution\cite{Atm4,SK4},
 and negligibly small mixing between the first
 and the third generations
 as $\sin^2 2 \theta_{13}\simeq 0$ 
 which is consistent with the CHOOZ experiment \cite{CHOOZ}.
Case (i) induces the maximal mixing between the first and the
second generations for the solar neutrino solution as
$\sin^2 2 \theta_{12}\simeq 1$\cite{S3a,S3b,demo-tani,O3}.
On the other hand,
case (ii) induces the small mixing between the first and the
second generations,
since the maximal mixing angles induced from
both $M_l$ and $M_\nu$ are canceled with each other\cite{S3b}.

\vspace{5mm}
%
%

Now let us estimate quantum corrections of
the MNS matrix in the democratic type of mass matrix.
We take the diagonal base of the
 charged lepton mass matrix at the
 high energy scale $\mh$, where the democratic textures are induced, 
 for the RGE analysis.
In this base the neutrino mass matrix in \eqsref{after_break_i}
and \eqvref{after_break_ii} are written by
$V_l^{T} M_\nu (m_h^{}) V_l^{} \simeq
F^T M_\nu (m_h^{}) F$,
where we use the approximation of $V_l^{}\simeq F$.
Then, the neutrino mass matrix at 
$\mz$ scale is given by
\begin{equation}
M_{\nu}\left(\mz\right) =
\dfrac{M_\nu\left(\mz\right)_{33}}{M_\nu\left(m_h^{}\right)_{33}}\;
R_G F^{T} M_\nu (m_h^{}) F R_G \,,
\end{equation}
where the matrix $R_G$ shows the renormalization
effects,
which is defined as
\begin{equation}
R_G \equiv
\left(\begin{array}{ccc}
1+\eta &     0  & 0 \\
0     & 1+\eta & 0 \\
0     &     0  & 1 \\
\end{array}\right)\,.
\eqlab{RG}
\end{equation}
The small parameter $\eta$ is given by
\begin{eqnarray}
\eqlab{def_eta}
\eta & \simeq & 1 -
{\exp
\left(-\dfrac{1}{16\pi^2}
\int^{\ln\left(m_h^{}\right)}_{\ln\left(\mz\right)}
y_\tau^2 dt
\right)} ,  \nonumber \\
& \simeq &
\dfrac{1}{8\pi^2}
\dfrac{m_\tau^2}{v^2}\left(1+\tan^2\beta\right)
\ln \left(\dfrac{m_h^{}}{\mz}\right)\, ,
\end{eqnarray}
where $y_\tau$ is the Yukawa coupling of $\tau$ 
and 
$v^2 \equiv {\langle H_u \rangle}^2 +
{\langle H_d \rangle}^2$. 
We neglect the Yukawa couplings
of $e$ and $\mu$ in \eqref{RG},
since those contributions to the RGE
are negligibly small comparing to that of $\tau$\cite{HO1}.
Therefore the first and the second generations
receive the same RGE corrections as in \eqref{RG}.
Now let us check 
whether the mixing angles receive significant changes
by quantum corrections or not in both
cases of (i) and (ii).

\vspace{5mm}

In the base of charged lepton democratic mass matrix,
$M_\nu (\mz)$ in case (i) is written as
\begin{eqnarray}
M_\nu^{(i)} (\mz)
& = & F R_G F^{T} M_\nu^{(i)}(m_h^{})F R_G F^{T}\,, \\
& \simeq & \bar{c_{\nu}} \left(\begin{array}{ccc}
1+\bar{r} &   \bar{r} & \bar{r} \\
  \bar{r} & 1+\bar{r}+\epsilon & \bar{r} \\
  \bar{r} &   \bar{r} & 1+\bar{r}+\delta
\end{array}\right)
+
2 \eta \bar{c_{\nu}}
\left(\begin{array}{ccc}
1 & 0 & 0 \\
0 & 1 & 0 \\
0 & 0 & 1
\end{array}\right)\,,
\label{34}
\end{eqnarray}
where
\begin{equation}
\bar{r} \equiv r -\dfrac{2}{3}\eta , \;\;\;\;\;\;\;\;\;
\bar{c_{\nu}} \equiv 
\dfrac{M_\nu\left(\mz\right)_{33}}{M_\nu\left(m_h^{}\right)_{33}}
c_{\nu} .
\label{13}
\end{equation}
Here we neglect the small parameters of order $\epsilon^2$,
$\epsilon \eta$, and $\epsilon \delta$.
Equation (\ref{34}) means that
the MNS matrix at $\mz$ scale is obtained only by
using $\bar{r}$ instead of $r$
in \eqref{after_break_i}.
This had been already shown in Ref.\cite{demo-tani}.
Therefore all we have to do for the RGE analyses of
the mixing angles
is to trace the change of $\bar{r}$.
Here we remind that the magnitude of
 $\eta$ is completely determined
 by the value of $\tan \beta$ and 
 the scale of $m_h^{}$\cite{HO1}, 
 then $\bar{r}$ is determined by Eq.(\ref{13}).  

Now let us show how the MNS matrix changes
as the change of $\bar{r}$. \\
\vspace{3mm}
\noindent
\underline{(i-a): 
$1 \gg \delta \gg \epsilon \gg \left|\bar{r}\right|$}\\ 
Neglecting the second order of small parameters
of $\delta, \bar{r}$, and $\epsilon$,
mass eigenvalues of $M_{\nu}^{(i)}(\mz)$ are give by
\begin{equation}
\label{ei_ia}
\bar{c_{\nu}}(1+\bar{r}+2\eta), \;\;\;\;\;
\bar{c_{\nu}}(1+\bar{r}+\epsilon +2\eta),\;\;\;\;\;
\bar{c_{\nu}}(1+\bar{r}+\delta +2\eta) \,.
\end{equation}
Then the unitary matrix $U_{\nu}$
becomes
\begin{equation}
U_{\nu} \simeq
\left(
    \begin{array}{ccc}
    1 & \dfrac{\bar{r}}{\epsilon} & \dfrac{\bar{r}}{\delta} \\
    -\dfrac{\bar{r}}{\epsilon} & 1 &\dfrac{\bar{r}}{\delta} \\
    -\dfrac{\bar{r}}{\delta} & -\dfrac{\bar{r}}{\delta} & 1
\end{array}
\right) ,
\end{equation}
which induces the MNS matrix $V_{MNS}$
as
\begin{equation}
V_{MNS} \simeq F^TU_{\nu} =
\left(
\begin{array}{ccc}
   \dfrac{1}{\sqrt{2}} \left( 1+\dfrac{\bar{r}}{\epsilon} \right)
& -\dfrac{1}{\sqrt{2}} \left( 1-\dfrac{\bar{r}}{\epsilon} \right)
& 0 \\
     \dfrac{1}{\sqrt{6}}
\left(1+2\dfrac{\bar{r}}{\delta}-\dfrac{\bar{r}}{\epsilon} \right)
&   \dfrac{1}{\sqrt{6}}
\left(1+2\dfrac{\bar{r}}{\delta}+\dfrac{\bar{r}}{\epsilon} \right)
& -\sqrt{\dfrac{2}{3}} \left( 1-\dfrac{\bar{r}}{\delta} \right) \\
   \dfrac{1}{\sqrt{3}}
\left( 1-\dfrac{\bar{r}}{\delta}-\dfrac{\bar{r}}{\epsilon} \right)
& \dfrac{1}{\sqrt{3}}
\left( 1-\dfrac{\bar{r}}{\delta}+\dfrac{\bar{r}}{\epsilon} \right)
& \dfrac{1}{\sqrt{3}} \left( 1+2\dfrac{\bar{r}}{\delta} \right) \\
\end{array}
\right) .
\end{equation}
Thus the mixing angles are given by
\begin{equation}
\sin^2 2\theta_{12} \simeq 1-4(\dfrac{\bar{r}}{\epsilon})^2, \;\;\;\;\;
\sin^2 2\theta_{13} \simeq 0, \;\;\;\;\;
\sin^2 2\theta_{23} \simeq \dfrac{8}{9}
\left(1+2\dfrac{\bar{r}}{\delta}
  -9(\dfrac{\bar{r}}{\delta})^2 \right) .
\end{equation}
This means all mixing angles are not changed
by quantum corrections
in the region of
$1 \gg \delta \gg \epsilon \gg | \bar{r}|$.
The value of $\sin ^2 2 \theta _{13}$ is negligible, 
since  $\sin ^2 2 \theta_{13} \simeq O(L_{21}^2) \ll 1$, 
where $L_{21}$ is the (21) component of L.
Equation (\ref{ei_ia}) shows that
$\Delta m_{12}^2 \simeq 2 \bar{c_{\nu}}^2 \epsilon$ and
$\Delta m_{23}^2 \simeq 2 \bar{c_{\nu}}^2 \delta$.
In order for the symmetry breaking parameter
$\delta$ to be smaller than symmetric terms of order
one, it must be
that $\delta \leq O(0.1)$.
On the other hand, neutrinoless $\beta \beta$-decay experiments 
\cite{double}
suggest $\bar{c_{\nu}}\leq O(0.1)$ eV.
Then, 
$\bar{c_{\nu}}= O(0.1)$ eV and $\delta = O(0.1)$
are obtained from the experimental results of
$\Delta m^2_{\rm ATM} \simeq 10^{-3}$ eV$^2$.
This means that
$\epsilon =O(10^{-3})$ for the MSW-L solution, and
$\epsilon =O(10^{-8})$ for the VO solution.
Then, the region of $\epsilon \gg \bar{r}$
corresponds to $\tan \beta < 10$\footnote{
The higher the scale of $m_h$ becomes, 
the smaller the value of $\tan \beta$
must be in order for the maximal flavor mixing not to be
destroyed by quantum corrections. 
} for the MSW-L solution
and $\tan \beta \ll 1$ for the VO solution\cite{HO1}.
Since the region of $\tan \beta \ll 1$ is excluded by
the Higgs search experiments\cite{date},
we can conclude the maximal mixing of
the VO solution in the democratic type of
mass matrix,
discussed in Refs.\cite{S3a,S3b},
is completely spoiled by
 quantum corrections\footnote{
The maximal mixing of the VO solution
is not spoiled by
 quantum corrections (even in
$\tan \beta = 3$), if
$m_h^{} \leq O(1)$ TeV.
However, such a low energy scale of $m_h^{}$
is not suitable from the view point of model building.
}. 
{}For the MSW-L solution, discussed 
 in Refs.\cite{demo-tani,O3},
 the sufficient condition of 
 $\tan \beta < 10$ must be satisfied.

\vspace{3mm}
\noindent
\underline{(i-b):
$1 \gg \delta \gg \left|\bar{r}\right| \gg \epsilon$}\\
Neglecting the second order of small parameters of
$\delta, \bar{r}$, and $\epsilon$,
mass eigenvalues of $M_{\nu}^{(i)}(\mz)$ are give by
\begin{equation}
\label{ei_ib}
\bar{c_\nu} (1+\dfrac{1}{2}\epsilon +2\eta) ,\;\;\;\;\;
\bar{c_\nu} (1+2\bar{r}+\dfrac{1}{2}\epsilon +2\eta) ,\;\;\;\;\;
\bar{c_\nu} (1+\bar{r}+\delta +2\eta) \,.
\end{equation}
The unitary matrix $U_{\nu}$ becomes
\begin{equation}
U_{\nu} \simeq
   \left(
\begin{array}{ccc}
        \dfrac{1}{\sqrt{2}} 
        \left( 1+\dfrac{1}{4}\dfrac{\epsilon}{\bar{r}}
\right)
      & \dfrac{1}{\sqrt{2}} 
      \left( 1-\dfrac{1}{4}\dfrac{\epsilon}{\bar{r}}
\right)
      & \dfrac{\bar{r}}{\delta} \\
        -\dfrac{1}{\sqrt{2}} 
        \left( 1-\dfrac{1}{4}\dfrac{\epsilon}{\bar{r}}
\right)
      & \dfrac{1}{\sqrt{2}} 
      \left( 1+\dfrac{1}{4}\dfrac{\epsilon}{\bar{r}}
\right)
      & \dfrac{\bar{r}}{\delta} \\
        -\dfrac{1}{2\sqrt{2}}\dfrac{\epsilon}{\delta}
      & -\sqrt{2}\dfrac{\bar{r}}{\delta}
      & 1
\end{array}
\right) ,
\end{equation}
which induces the MNS matrix as
\begin{equation}
V_{MNS} \simeq F^TU_{\nu}=
\left(
\begin{array}{ccc}
        1
      & -\dfrac{1}{4} \dfrac{\epsilon}{\bar{r}}
      & 0 \\
              \dfrac{1}{2\sqrt{3}} \dfrac{\bar{r}}{\delta}
      & \dfrac{1}{\sqrt{3}} 
      \left( 1+\dfrac{1}{2} \dfrac{\bar{r}}{\delta}
\right)
      & -\sqrt{\dfrac{2}{3}} 
      \left( 1-\dfrac{\bar{r}}{\delta} \right)
\\
        -\dfrac{1}{2\sqrt{6}} \left( \dfrac{\epsilon}{\delta}
        -\dfrac{\epsilon}{\bar{r}} \right)
      & \sqrt{\dfrac{2}{3}} 
        \left( 1-\dfrac{\bar{r}}{\delta} \right)
      & \dfrac{1}{\sqrt{3}} 
        \left( 1+2\dfrac{\bar{r}}{\delta} \right)
\\
\end{array}
\right)\,.
\end{equation}
Then the mixing angles are given by
\begin{equation}
\sin^2 2\theta_{12} \simeq \dfrac{1}{4} 
\left(\dfrac{\epsilon}{\bar{r}}\right)^2 ,
\;\;\;\;\;
\sin^2 2\theta_{13} \simeq 0 ,  \;\;\;\;\;
\sin^2 2\theta_{23} \simeq \dfrac{8}{9}
\left(1+2\dfrac{\bar{r}}{\delta}
    -9(\dfrac{\bar{r}}{\delta})^2 \right) .
\label{11}
\end{equation}
The value of $\sin ^2 2 \theta _{13}$ is negligible, 
since  $\sin ^22 \theta _{13} \simeq O(L_{21}^2) \ll 1$.
This means that maximal mixings 
of
all solar solutions in the democratic type of mass matrix
of Eq.(\ref{34}) are spoiled by quantum corrections
in the region
 of $1 \gg \delta \gg | \bar{r}| \gg \epsilon$,
although the mixings between the first and the third generations,
and between the second and the third generations
are stable against
quantum corrections \footnote{
If $L_{21} \simeq O(1)$, the mixing angle of $\sin 2\theta_{12}$ can be
stable agaist quantum corrections. 
However, it is difficult to obtain $L_{21} \simeq O(1)$ from $M_l^{(c)}$
} .

\vspace{3mm}
\noindent
\underline{(i-c): 
$1 \gg\ \left|\bar{r}\right| \gg \delta \gg \epsilon$}\\
Neglecting the second order of small parameters of
$\delta, \bar{r}$, and $\epsilon$,
 mass eigenvalues of $M_{\nu}^{(i)}(\mz)$ are give by
\begin{equation}
\label{ei_ic}
\bar{c_{\nu}}( 1+\dfrac{1}{2}\epsilon +2\eta),\;\;\;\;\;
\bar{c_{\nu}}
( 1+\dfrac{2}{3}\delta+\dfrac{1}{6}\epsilon +2\eta),\;\;\;\;\;
\bar{c_{\nu}}
( 1+3\bar{r}+\dfrac{1}{3}\delta+\dfrac{1}{3}\epsilon +2\eta)\,.
\end{equation}
In this case $U_{\nu}$ becomes
\begin{equation}
U_\nu \simeq
\left(
\begin{array}{ccc}
   \dfrac{1}{\sqrt{2}}
     \left( 1
        +\dfrac{1}{4}\dfrac{\epsilon}{\delta}
           +\dfrac{1}{6}\dfrac{\epsilon}{\bar{r}}
     \right)
& \dfrac{1}{\sqrt{6}}
     \left(1
      -\dfrac{3}{4}\dfrac{\epsilon}{\delta}
         +\dfrac{2}{9}\dfrac{\delta}{\bar{r}}
             -\dfrac{1}{9}\dfrac{\epsilon}{\bar{r}}
     \right)
& \dfrac{1}{\sqrt{3}}
     \left( 1
      -\dfrac{1}{9}\dfrac{\delta}{\bar{r}}
             -\dfrac{1}{9}\dfrac{\epsilon}{\bar{r}}
             \right) \\
   -\dfrac{1}{\sqrt{2}}
     \left(1
        -\dfrac{1}{4}\dfrac{\epsilon}{\delta}
           -\dfrac{1}{6}\dfrac{\epsilon}{\bar{r}}
           \right)
& \dfrac{1}{\sqrt{6}}
     \left(1
      +\dfrac{3}{4}\dfrac{\epsilon}{\delta}
         +\dfrac{2}{9}\dfrac{\delta}{\bar{r}}
             -\dfrac{1}{9}\dfrac{\epsilon}{\bar{r}}
             \right)
& \dfrac{1}{\sqrt{3}}
     \left(1
      -\dfrac{1}{9}\dfrac{\delta}{\bar{r}}
             +\dfrac{2}{9}\dfrac{\epsilon}{\bar{r}}
             \right) \\
        -\dfrac{1}{2\sqrt{2}}
        \left( \dfrac{\epsilon}{\delta}
           -\dfrac{1}{3}\dfrac{\epsilon}{\bar{r}}
           \right)
& -\sqrt{\dfrac{2}{3}}
        \left(1
         -\dfrac{1}{9}\dfrac{\delta}{\bar{r}}
             +\dfrac{1}{18}\dfrac{\epsilon}{\bar{r}}
             \right)
& \dfrac{1}{\sqrt{3}}
     \left(1
      +\dfrac{2}{9}\dfrac{\delta}{\bar{r}}
             -\dfrac{1}{9}\dfrac{\epsilon}{\bar{r}}
             \right) \\
\end{array}
\right)\,.
\end{equation}
Thus, the MNS matrix is given by
\begin{equation}
V_{MNS} \simeq F^TU_\nu=
\left( 
\begin{array}{ccc}
   1 
& -\dfrac{\sqrt{3}}{4}\dfrac{\epsilon}{\delta} 
& -\dfrac{1}{3\sqrt{6}}\dfrac{\epsilon}{\bar{r}} \\
   \dfrac{\sqrt{3}}{4}\dfrac{\epsilon}{\delta}
& 1
& -\dfrac{\sqrt{2}}{9} \left( \dfrac{\delta}
{\bar{r}}-\dfrac{1}{2}\dfrac{\epsilon}{\bar{r}} \right) \\
   \dfrac{1}{2\sqrt{6}}\dfrac{\epsilon}{\bar{r}}
& \dfrac{\sqrt{2}}{9} \left(\dfrac{\delta}
{\bar{r}}-\dfrac{1}{2}\dfrac{\epsilon}{\bar{r}} \right)
& 1 \\
\end{array}
\right)\,,
\end{equation}
which gives the mixing angles as 
\begin{equation} 
\sin^2 2\theta_{12} \simeq \dfrac{3}{4} 
\left(\dfrac{\epsilon}{\delta}\right)^2 ,  \;\;\;\;\; 
\sin^2 2\theta_{13} \simeq \dfrac{1}{54} 
\left(\dfrac{\epsilon}{\bar{r}}\right)^2 ,  \;\;\;\;\; 
\sin^2 2\theta_{23} \simeq \dfrac{8}{81} 
\left(\dfrac{\delta}{\bar{r}}
  +\dfrac{\epsilon}{4\bar{r}} \right)^2 .
\label{12}
\end{equation}
This means that large mixing angles in 
both the solar and the atmospheric 
neutrino solutions are spoiled 
by quantum corrections
in the region of $1 \gg\ | \bar{r}| \gg \delta \gg \epsilon$. 
It is because 
the condition of $| \bar{r}| \gg \delta , \epsilon$
induces 
$U_{\nu} \simeq F$, 
which is just the case of Eq.(\ref{FF}).

\par
The conclusion in case (i) are that 
(1): the maximal mixing of the
VO solution is
destroyed by quantum corrections, and 
(2): the sufficient condition of
$\tan \beta < 10$ must be satisfied for
the MSW-L solution.

\vspace{5mm}

Next, let us show the case (ii) 
in the base of charged lepton democratic mass matrix, 
where $M_\nu^{(ii)} (\mz)$ is written by
\begin{equation}
M_\nu ^{(ii)}(\mz) \simeq \bar{c_{\nu}}
\left(\begin{array}{ccc}
1+\bar{r} &   \bar{r}+\epsilon & \bar{r} \\
  \bar{r}+\epsilon & 1+\bar{r} & \bar{r} \\
  \bar{r} &   \bar{r} & 1+\bar{r}+\delta
\end{array}\right)
+
2 \eta \bar{c_{\nu}}
\left(\begin{array}{ccc}
1 & 0 & 0 \\
0 & 1 & 0 \\
0 & 0 & 1
\end{array}\right)\,.
\label{35}
\end{equation}
Let us show how the MNS matrix changes 
according to the change of $\bar{r}$
as in case (i). 

\vspace{3mm}
\noindent
\underline{(ii-a): $1 \gg \delta \gg \epsilon , | \bar{r}|$}\\ 
Neglecting the second order of small parameters of
$\delta, \bar{r}$, and $\epsilon$,
the mass eigenvalues of $M_{\nu}^{(ii)}(m_Z)$ 
are give by
\begin{equation}
\label{ei_iia}
\bar{c_{\nu}}( 1-\epsilon +2\eta),\;\;\;\;\;
\bar{c_{\nu}}( 1+2\bar{r}+\epsilon +2\eta), \;\;\;\;\;
\bar{c_{\nu}}( 1+\bar{r}+\delta +2\eta).
\end{equation}
In this case
$U_{\nu}$ becomes
\begin{equation}
U_{\nu} \simeq
   \left(
\begin{array}{ccc}

        \dfrac{1}{\sqrt{2}}
      & \dfrac{1}{\sqrt{2}}
      & \dfrac{\bar{r}}{\delta} \\
        -\dfrac{1}{\sqrt{2}}
      & \dfrac{1}{\sqrt{2}}
      & \dfrac{\bar{r}}{\delta} \\
        0
      & -\sqrt{2}\frac{\bar{r}}{\delta}
      & 1
\end{array}
\right) ,
\end{equation}
which induces
the MNS matrix as
\begin{eqnarray}
& & V_{MNS} = L^{\dagger}F^TU_{\nu} \\
& & \simeq \left(
\begin{array}{ccc}
        1
      & \dfrac{1}{\sqrt{3}}L_{21} 
          \left( 1+2\dfrac{\bar{r}}{\delta} \right)
      & -\sqrt{\dfrac{2}{3}}L_{21}
          \left(1-\dfrac{\bar{r}}{\delta} \right)\\
              L_{12}
      & \dfrac{1}{\sqrt{3}} 
          \left( 1+2 \dfrac{\bar{r}}{\delta} \right)
        +\dfrac{1}{\sqrt{3}}L_{31} 
          \left( 1+\dfrac{\bar{r}}{\delta} \right)
      & -\sqrt{\dfrac{2}{3}} 
          \left( 1- \dfrac{\bar{r}}{\delta} \right)
        +\dfrac{1}{\sqrt{3}}L_{32} 
          \left( 1+2\dfrac{\bar{r}}{\delta} \right)
\\
        L_{13}
      & \sqrt{\dfrac{2}{3}} 
          \left(1-\dfrac{\bar{r}}{\delta} \right)
        +\dfrac{1}{\sqrt{3}}L_{23} 
          \left( 1+2 \dfrac{\bar{r}}{\delta}
\right)
      & \dfrac{1}{\sqrt{3}}
          \left(1+2\dfrac{\bar{r}}{\delta} \right)
        -\dfrac{3}{2}L_{23}
          \left(1-\dfrac{\bar{r}}{\delta}\right)
\\
\end{array}
\right) , \nonumber
\end{eqnarray}
where we revive the small elements
of $L_{ij}$ $(i \neq j)$.
This shows that the mixing angles are given by
\begin{equation}
\sin^2 2\theta_{12} \simeq \dfrac{4}{3}L_{21}^2
      \left( 1+2\dfrac{\bar{r}}{\delta} \right) ^2 ,  \;\;\;
\sin^2 2\theta_{13} \simeq \dfrac{8}{3}L_{21}^2
      \left( 1-\dfrac{\bar{r}}{\delta} \right) ^2 ,  \;\;\;
\sin^2 2\theta_{23} \simeq \dfrac{8}{9}
\left(1+2\dfrac{\bar{r}}{\delta}-9(\dfrac{\bar{r}}{\delta})^2 \right) .
\label{21}
\end{equation}
This means that all flavor mixings
are not spoiled by quantum corrections
in the region of $1 \gg \delta \gg | \bar{r}| , \epsilon$.
Equation (\ref{ei_iia}) suggests that
$\Delta m_{12}^2 
      \simeq 4 \bar{c_{\nu}}^2 ( \bar{r}+ \epsilon )$ and
$\Delta m_{23}^2 
      \simeq 2 \bar{c_{\nu}}^2 \delta$.
Thus, when $| \bar{r} | \geq \epsilon$,
quantum correction
is the origin of mass squared difference
for the solar neutrino solution.
Where we must tune the value of $\tan \beta$
in order to obtain the suitable mass squared difference.
The case of
$\tan \beta \simeq 10$
induces $\Delta m_{12}^2 \sim 10^{-5}$ eV$^2$
at $m_Z$.
Therefore the condition of $\tan \beta \leq 10$
must be satisfied
in order to obtain the suitable magnitude of
mass squared difference,
for the MSW-S solution discussed in Ref.\cite{S3b}.
As for the mixings between the first and the third
generations, and between the second and the third
generations, Eq.(\ref{21}) shows
that they are stable against
quantum corrections.

\vspace{3mm}
\noindent
\underline{(ii-b): 
$1 \gg\ | \bar{r} | \gg \delta \gg \epsilon$} \\
Neglecting the second order of small parameters of
$\delta, \bar{r}$, and $\epsilon$,
 mass eigenvalues of $M_{\nu}^{(ii)}(m_Z)$ are give by
\begin{equation}
\label{ei_iic}
\bar{c_{\nu}}
 ( 1-\epsilon +2\eta),\;\;\;\;\;
\bar{c_{\nu}}
 ( 1+\dfrac{2}{3}\delta+\dfrac{1}{3}\epsilon +2\eta),\;\;\;\;\;
\bar{c_{\nu}}
 ( 1+3\bar{r}+\dfrac{1}{3}\delta+\dfrac{2}{3}\epsilon +2\eta) .\;\;\;\;\;
\end{equation}
In this case
$U_{\nu}$ becomes
\begin{equation}
U_{\nu} \simeq
\left(
\begin{array}{ccc}
   \dfrac{1}{\sqrt{2}}
& \dfrac{1}{\sqrt{6}}
      \left(1
      +\dfrac{2}{9}\dfrac{\delta}{\bar{r}}
             -\dfrac{2}{9}\dfrac{\epsilon}{\bar{r}}
             \right)
& \dfrac{1}{\sqrt{3}}
      \left(1
      -\dfrac{1}{9}\dfrac{\delta}{\bar{r}}
             -\dfrac{2}{9}\dfrac{\epsilon}{\bar{r}}
             \right) \\
   -\dfrac{1}{\sqrt{2}}
&  \dfrac{1}{\sqrt{6}}
     \left(1
         +\dfrac{2}{9}\dfrac{\delta}{\bar{r}}
             -\dfrac{2}{9}\dfrac{\epsilon}{\bar{r}}
             \right)
& \dfrac{1}{\sqrt{3}}
    \left(1
      -\dfrac{1}{9}\dfrac{\delta}{\bar{r}}
             -\dfrac{2}{9}\dfrac{\epsilon}{\bar{r}}
             \right) \\
        0
& -\sqrt{\dfrac{2}{3}}
    \left(1
         -\dfrac{1}{9}\dfrac{\delta}{\bar{r}}
             +\dfrac{1}{9}\dfrac{\epsilon}{\bar{r}}
             \right)
& \dfrac{1}{\sqrt{3}}
    \left(1
      +\dfrac{2}{9}\dfrac{\delta}{\bar{r}}
             +\dfrac{4}{9}\dfrac{\epsilon}{\bar{r}}
             \right)\\
\end{array}
\right) ,
\end{equation}
which induces the MNS matrix as
\begin{equation}
V_{MNS} =L^{\dagger} F^T U_{\nu} =
\left( 
\begin{array}{ccc}
   1 & L_{21}
& -\dfrac{\sqrt{2}}{9} L_{21}
      \left( \dfrac{\delta}{\bar{r}}
+\dfrac{\epsilon}{\bar{r}}\right) \\
   L_{12}
& 1+L_{32}\left( \dfrac{\delta}{\bar{r}}
-\dfrac{\epsilon}{\bar{r}}\right)
& -\dfrac{\sqrt{2}}{9}
      \left( \dfrac{\delta}{\bar{r}}
-2\dfrac{\epsilon}{\bar{r}} \right)
+L_{32}\\
   L_{13}
& \dfrac{\sqrt{2}}{9}
     \left( \dfrac{\delta}{\bar{r}}
-\dfrac{\epsilon}{\bar{r}} \right) +L_{23}
& 1- \dfrac{\sqrt{2}}{9}
     \left( \dfrac{\delta}{\bar{r}}
-\dfrac{\epsilon}{\bar{r}} \right)\\
\end{array}
\right) .
\end{equation}
This suggests that 
the mixing angles are given by
\begin{equation}
\sin^2 2\theta_{12} \simeq 4L_{21}^2 ,  \;\;\;\;\;
\sin^2 2\theta_{13} \simeq \dfrac{2}{81}L_{21}^2
                    \left( \dfrac{\delta}{\bar{r}}
                    +\dfrac{\epsilon}{\bar{r}}
                    \right) ^2 ,  \;\;\;\;\;
\sin^2 2\theta_{23} \simeq \frac{8}{81}
\left(\frac{\delta}{\bar{r}}
+2\frac{\epsilon}{\bar{r}} \right)^2 ,
\label{22}
\end{equation}
which means that the large mixing of
the atmospheric neutrino solution
is destroyed in the region of
$1 \gg\ | \bar{r}| \gg \delta \gg \epsilon$.
It is because
the condition of $| \bar{r}| \gg \delta , \epsilon$
induces $U_{\nu} \simeq F$, which is just
the case of Eq.(\ref{FF}).

\par
The conclusion in case (ii) is that 
sufficient condition of
$\tan \beta \leq 10$ must be satisfied for
the MSW-S solution.

\vspace{5mm}

The democratic type of mass matrix texture
is one of the most interesting
candidate of quark and lepton mass matrices,
which has been said to be able to
induce the suitable
solutions of the atmospheric and the solar neutrino problems.
In this paper,
we investigate whether the lepton
flavor mixing angles in the
democratic type of
mass matrix are stable against quantum corrections or not
in the minimal supersymmetric standard model 
with the dimension five
operator which induces the neutrino Majorana mass matrix.
We take the simple breaking patterns 
of $S_3{}_L \times S_3{}_R$ or
$O(3)_L \times O(3)_R$ symmetries, and
the scale where democratic textures are induced as
$O(10^{13})$ GeV.
Under the above conditions,
we find that the stability of mixing angles
in the democratic type of mass matrix
against quantum corrections
depends on the solar neutrino solutions.
The maximal mixing of the VO solution is spoiled
by quantum corrections 
 in the experimentally allowed region of 
 $\tan \beta$. 
The MSW-L solution is spoiled by
quantum corrections in the region of $\tan \beta > 10$.
On the other hand, the condition of
$\tan \beta \leq 10$ is needed in order to
obtain the suitable mass squared difference of
the MSW-S solution.
These strong constraints must be regarded
for the model building of the democratic
type of mass matrix.
If we take $m_h^{}$ as the GUT scale of $O(10^{16})$ GeV,
the constraints for the stability of the mixing angles
against quantum corrections 
become more severe, that is, 
the MSW-L solution needs 
$\tan \beta < 8$, 
and the MSW-S solution needs
$\tan \beta \leq 8$.


\vspace{5mm}
We would like to thank T. Yanagida and M. Bando
for the suggestion of this work.
The work of NO is supported by the JSPS Research Fellowship
for Young Scientists, No.2996.

%
%

\end{document}